\documentclass[conference]{IEEEtran}
\IEEEoverridecommandlockouts
\usepackage{cite}
\usepackage{amsmath,amssymb,amsfonts}
\usepackage{algorithmic}
\usepackage{siunitx}
\usepackage{graphicx}
\usepackage{textcomp}
\usepackage{xcolor}
\def\BibTeX{{\rm B\kern-.05em{\sc i\kern-.025em b}\kern-.08em
    T\kern-.1667em\lower.7ex\hbox{E}\kern-.125emX}}
\begin{document}

\title{A CubeSat platform for space based quantum key distribution
\thanks{This research is supported by the National Research Foundation, Singapore, under its Central Gap Fund (NRF2018NRFCG001-001).}
}

\author{\IEEEauthorblockN{Srihari Sivasankaran}
\IEEEauthorblockA{\textit{Centre for Quantum Technologies} \\
Singapore 117543\\
cqtsri@nus.edu.sg}
\and
\IEEEauthorblockN{Clarence Liu}
\IEEEauthorblockA{\textit{Centre for Quantum Technologies} \\
Singapore 117543 \\
clarenceliuh@nus.edu.sg}
\and
\IEEEauthorblockN{Moritz Mihm}
\IEEEauthorblockA{\textit{Centre for Quantum Technologies} \\
Singapore 117543 \\
mmihm@nus.edu.sg}
\and
\IEEEauthorblockN{Alexander Ling}
\IEEEauthorblockA{\textit{Centre for Quantum Technologies} \\
Singapore 117543 \\
alexander.ling@nus.edu.sg}
}

\maketitle

\begin{abstract}
Satellite nodes are an enabler of global quantum networks by overcoming the distance limitations of fiber and free-space links on ground. The design of quantum sources and receivers for satellites, however, is challenging in terms of size, weight, and power consumption, as well as mechanical and thermal stability. This is all the more true for cost-efficient nanosatellites such as the popular CubeSat platform standard.

Here we report on the follow-up mission of SpooQy-1, a 3U CubeSat that successfully demonstrated the generation of polarization-entangled photons in orbit. The next iteration of the mission will showcase satellite-to-ground quantum key distribution based on a compact source of polarization-entangled photon-pairs, and we have recently completed the integration of a fully functional demonstrator as a milestone towards the flight model.

We also briefly describe the design of the optical ground station that we are currently building in Singapore for receiving the quantum signal. We present the most important subsystems and illustrate the concept of operation.
\end{abstract}


\section{Introduction}
Quantum key distribution (QKD) has reached a level of maturity that makes it possible to start building global networks. In this effort, network nodes in space are helping to overcome terrestrial distance limitations, and there are numerous initiatives underway to do so \cite{a1, a2}. A pathfinder in that direction was SpooQy-1, a 3U CubeSat that demonstrated the generation of entangled photon pairs in orbit \cite{a3}.

Here we report on the follow-up mission of SpooQy-1 and the corresponding receiver station on the ground. The concept of operation is for the CubeSat's science instrument to generate entangled photon pairs and send one of the photons to the ground station receiver. The underlying protocol is BBM92 \cite{a4}, an entanglement version of the BB84 protocol \cite{a5}.

We have recently completed the assembly and integration of a fully functional demonstrator of said science instrument. Our compact source generates polarization-entangled photon pairs by spontaneous parametric downconversion (SPDC) based on a periodically poled potassium titanyl phosphate crystal \cite{a6}. One photon of each pair is analyzed in the science instrument, whereas the other photon is sent to a receiver on the ground. We are in the process of constructing such an optical ground station (OGS) for the reception and analysis of polarization-entangled photons in Singapore. The system comprises a telescope, pointing, acquisition, and tracking (PAT) system, as well as a polarization correction system (PCS) and quantum receiver.

We present the architecture of the science instrument in section \ref{sec:payload_design} before discussing the performance in section \ref{sec:performance}. The high-level design and concept of operations of the ground station and respective subsystems are presented in section \ref{sec:ogs}. The manuscript concludes with a summary in section \ref{sec:conclusion}.

\section{Science instrument architecture}
\label{sec:payload_design}
\begin{figure*}[htbp]
    \centerline{\includegraphics{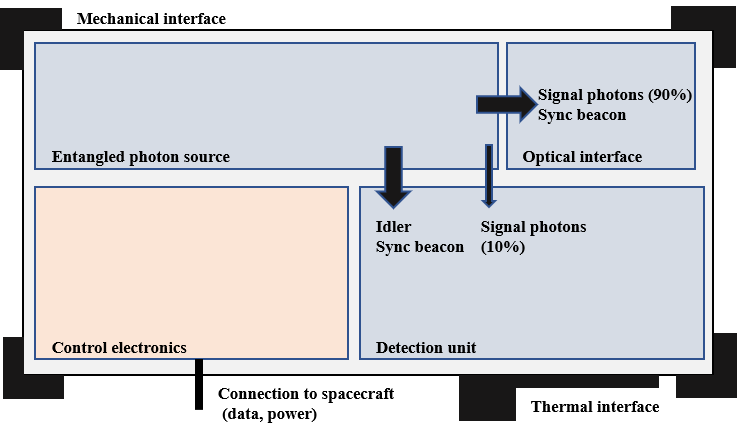}}
    \caption{System architecture of the science instrument. Adapted from \cite{a7}.}
    \label{fig:architecture}
\end{figure*}

Fig.~\ref{fig:architecture} shows the system architecture of the science instrument, which includes the entangled photon source, a detection unit, control electronics, and an optical ground station interface.

The source produces photon pairs of different wavelengths (non-degenerate), entangled in the polarization degree of freedom. While one photon of each pair (the idler at \SI{837}{nm}) is measured in the detection unit within the science instrument, the other photon (the signal at \SI{785}{nm}) together with a timing beacon is sent to ground via the optical ground station interface. A small fraction of the signal photons, however, is analyzed in the self-check unit of the science instrument's detection system for entanglement quality and internal efficiency.

A dichroic mirror separates idler and signal photons for detection on board and on ground, respectively. While all idler photons are analyzed within the science instrument, the signal photons (and beacon) are split using a non-polarizing beam splitter, guiding about \SI{90}{\%}\footnote{While the presented instrument uses a 90/10 beam splitter, the intended design was for a 99/1 beam splitter.} towards the optical ground station interface. The remaining fraction of the signal photons are analyzed in the self-check unit.

\begin{figure*}[htbp]
    \centerline{\includegraphics{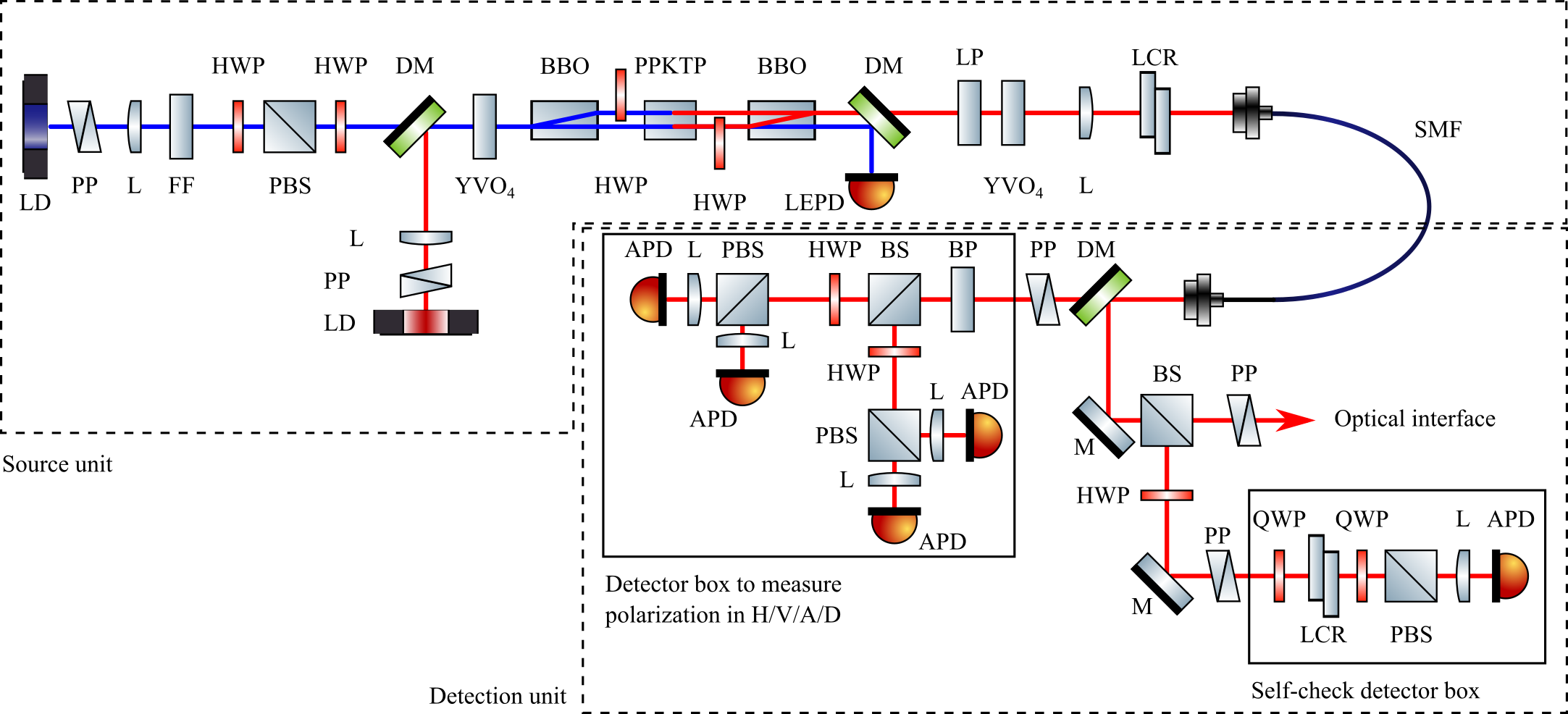}}
    \caption{Schematic of the science instrument. LD: laser diode; PP: prism pair; L: lens; FF: fluorescence filter; HWP: half-wave plate; PBS: polarizing beam splitter; DM: dichroic mirror; YVO$_4$: yttrium orthovanadate; BBO: beta barium borate; PPKTP: periodically poled potassium titanyl phosphate; LEPD: lateral effect photodiode; LP: long pass filter; LCR: liquid crystal retarder; SMF: single-mode fiber; BP: band pass filter; BS: beam splitter; APD: avalanche photodiode; M: mirror; QWP: quarter-wave plate.}
    \label{fig:schematic}
\end{figure*}

The science instrument is implemented as a series of components, mounted onto a baseplate. To comply with the CubeSat form factor, the baseplate has a size of {\SI{200}{mm} x \SI{200}{mm}}, while the total mass and electrical power consumption are estimated at \SI{3650}{\gram} and \SI{30}{\watt}, respectively. Fig.~\ref{fig:schematic} shows a schematic of the science instrument, which is explained in further detail below.

\subsection{Source unit}
The entangled states are created in a beam displacement interferometer in which two orthogonally polarized pump beams are downconverted in the same periodically poled crystal, and then recombined \cite{a6}.

Following the beam path, the pump beam at \SI{405}{nm} passes a prism pair for beam alignment, a lens for focusing into the periodically poled crystal, a filter for removal of fluorescence, a unit of half-wave plate and polarizing beam splitter for intensity adjustment, and another half-wave plate to rotate the polarization to \SI{45}{\degree} before entering the interferometer.

The horizontal and vertical projections of the diagonally polarized pump beam are spatially separated in a beta barium borate (BBO) crystal. The horizontally polarized beam is then converted to vertical polarization by means of a half-wave plate and both beams are downconverted in a temperature-stabilized periodically poled potassium titanyl phosphate (PPKTP) crystal. The daughter photons with wavelengths of \SI{837}{nm} (idler) and \SI{785}{nm} (signal) have the same vertical polarization as the pump photons (SPDC Type-0). The polarization of one beam is converted to horizontal before both beams are recombined in a second BBO crystal. The remaining pump light is removed by a dichroic mirror and long pass filter and the downconverted beam is collimated by a second lens.

To compensate for phase differences of signal and idler due to wavelength dependencies and different path lengths in the nonlinear crystals, we use two yttrium orthovanadate (YVO$_4$) crystals sandwiching the interferometer. Additionally, we use a liquid crystal retarder (LCR), placed behind the collimating lens in the source unit, to actively control the phase between signal and idler.

For time synchronization of detector events in the receivers on board and on ground, a pulsed laser at \SI{785}{nm} with tunable frequency between \SI{1}{kHz} and \SI{50}{kHz} and pulse width of \SI{5}{ns} is used. The beacon beam is superimposed with the pump beam using a dichroic mirror and follows the same beam path as signal and idler, eventually being coupled into a single-mode fiber for spatial mode filtering and guidance to the detection unit.

\subsection{Detection unit}
On the detection side, signal and idler are first separated by a dichroic mirror for detection on board and guidance to a ground station receiver, respectively (see Fig.~\ref{fig:schematic}). The idler beam is transmitted through the dichroic mirror and, upon alignment with a prism pair and filtering, the polarization is measured in the four standard channels horizontal (H), vertical (V), anti-diagonal (A), and diagonal (D). A 50/50 beam splitter randomly selects the measurement basis (H/V or A/D), a combination of half-wave plate and polarizing beam splitter projects the photons on the respective states. Measurement is achieved by focusing the light on avalanche photodiodes (APDs).

The signal beam, reflected at the dichroic mirror, is first split for guidance to a ground station receiver and analysis on board. While around \SI{90}{\%} of the signal photons are sent to ground, the remaining fraction is sent to a self-check detector unit, which is used to measure the entanglement quality and efficiencies.

\newpage

\section{Performance}
\label{sec:performance}
The figure of merit of space based quantum key distribution is the transmitted key rate, in our case from the source on board a CubeSat to a ground station receiver. Considering only the entangled photon source, the relevant parameters are brightness (rate of generated photon pairs per second) and entanglement quality or visibility $\mathrm{VIS}$, which is related to the quantum bit error rate (QBER): $\mathrm{QBER}=(1-\mathrm{VIS})/2$.

We have developed a model to study the key rates in different scenarios and we find that maximum key rate requires a source brightness of at least $\SI{25e6}{counts/sec}$ and a total QBER of less than \SI{5}{\%} on average \cite{a7, a8}. The target source QBER is \SI{1}{\%}, corresponding to a visibility of \SI{98}{\%}.


Our source performance values are summarized in Table~\ref{tab:performance}. We measure a source brightness of \SI{13.6E6}{counts\per\second\per\milli\watt}. Thus we can estimate a minimum pump power of \SI{1.84}{mW} is required to meet the target generated pair rate.

\begin{table}[htbp]
    \caption{Source performance overview}
    \begin{center}
    \begin{tabular}{|c|c|}\hline
        \textbf{Parameter}  & \textbf{Value}\\\hline
        Pump current       & \SI{47}{\milli\ampere}\\\hline
        Pump temperature   & \SI{31.50}{\degreeCelsius}\\\hline
        PPKTP temperature  & \SI{28.25}{\degreeCelsius}\\\hline
        Brightness         & \SI{13.6E6}{counts\per\second\per\milli\watt}\\\hline
    \end{tabular}
    \label{tab:performance}
    \end{center}
\end{table}

\begin{figure}[htbp]
    \centerline{\includegraphics[width=1\linewidth,height=6cm]{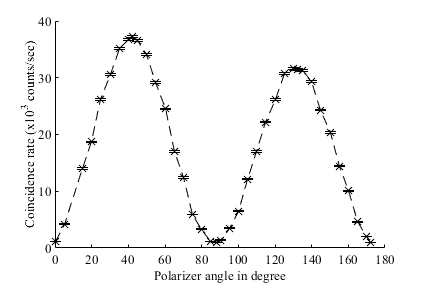}}
    \caption{Single polarizer visibility measurement of the polarization entangled photon pairs \cite{a9}.}
    \label{fig:visibility}
\end{figure}

For the source visibility (single polarizer measurement\footnote{The single polarizer measurement is useful for a quick characterization of entangled photon pair sources \cite{a9}.}), see Fig.~\ref{fig:visibility}. The visibility curve corresponds to the vertical (VV) and horizontal (HH) projected states at maxima, while minima correspond to diagonal (DD) and anti-diagonal (AA) polarization states. The difference in the peak maxima indicates an intensity imbalance between both arms of the interferometer.

Based on the maximum and minimum coincidence counts $c_\mathrm{max}$ and $c_\mathrm{min}$ of Fig.~\ref{fig:visibility}, the visibility is:
\begin{equation}
    \mathrm{VIS}=\frac{c_\mathrm{max}-c_\mathrm{min}}{c_\mathrm{max}+c_\mathrm{min}}=\frac{{38}-{1}}{{38}+{1}}=\SI{94.9}{\%}\,.
\end{equation}
The value indicates that the phase compensation between signal and idler is not completely optimized at the stage of recording.

\section{Optical ground station}
\label{sec:ogs}
\begin{figure}[htbp]
    \centerline{\includegraphics[width=1\linewidth]{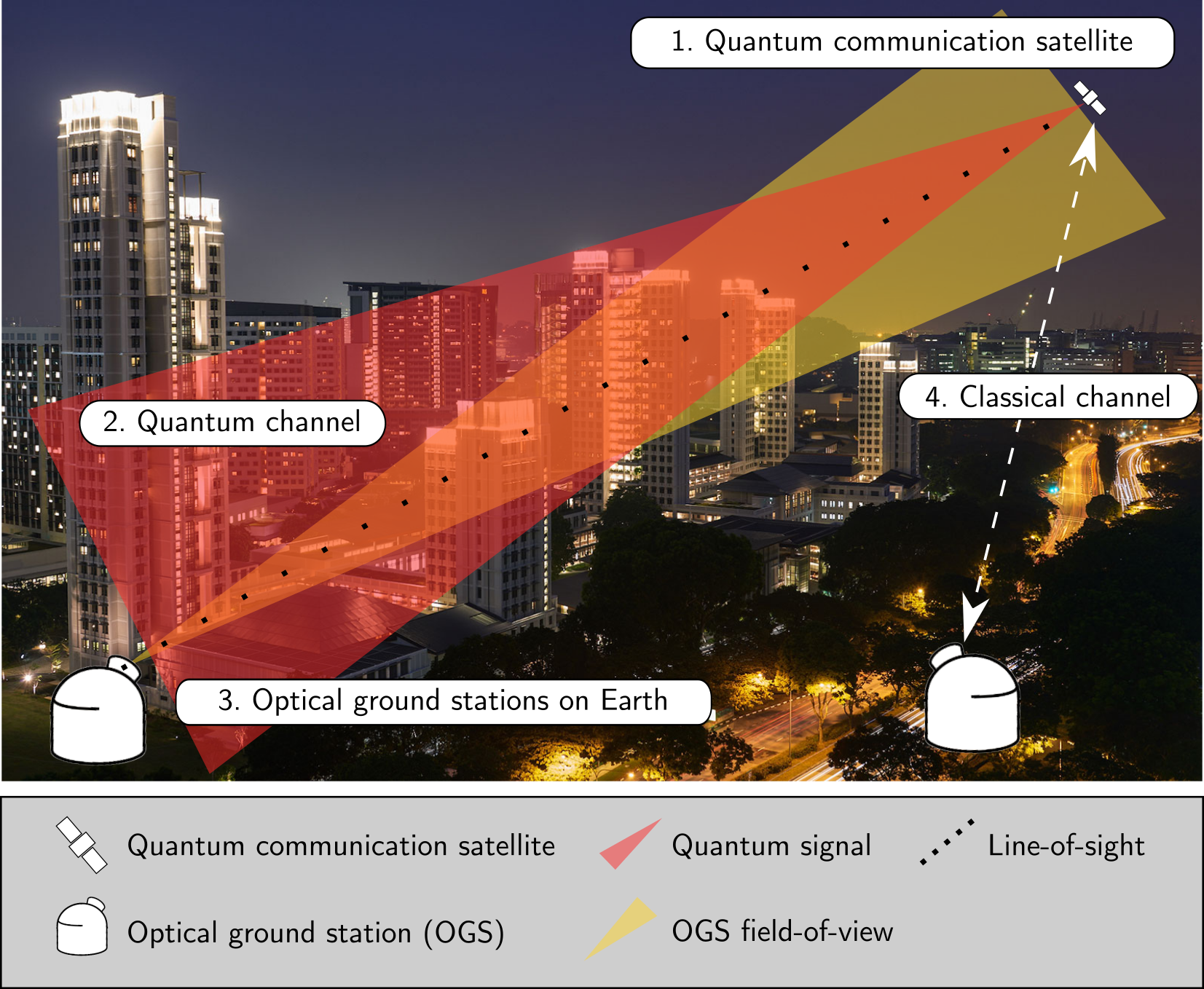}}
    \caption{Overview of the key components for space based quantum key distribution.}
    \label{fig:ogsintro}
\end{figure}

The basic components for space based quantum key distribution are: the quantum communication satellite, a quantum channel, optical ground station(s), and a classical channel (see Fig.~\ref{fig:ogsintro}). During a satellite pass, the satellite and OGS needs to establish a line-of-sight for communication to take place. The quantum channel refers to the transmission of the signal photons generated on the satellite through the Earth's atmosphere along the line-of-sight, while the classical channel allows and establishes a high bandwidth communication channel for post-processing of the communicated data. In our case, the satellite is in a low Earth orbit (LEO) and the OGS is installed on the campus of the National University of Singapore. We describe some of the key OGS subsystems and considerations for receiving the quantum signal in this section.

The primary goal of the OGS is to establish an optical link with the satellite, and to detect and analyze the quantum signal sent down to Earth. To achieve this goal, the OGS requires several critical subsystems -- a telescope system (optical receiver and motorised tracking mount); PAT system including uplink beacon, coarse and fine tracking system; PCS; and a quantum receiver. A motorised tracking mount provides the optical receiver with the basic capability of following a satellite pass. To establish a good quality line-of-sight, the OGS is also equipped with a PAT system which utilises a polarized downlink beacon that is sent down from the satellite for pointing and tracking. Simultaneously, an uplink beacon helps the satellite to locate and orientate itself at the OGS to establish a line-of-sight. Together, the PAT system works to keep the satellite within the field-of-view of the quantum receiver (QFOV) which detects and analyses the quantum signal. We have deliberately restricted the QFOV to approximately \SI{15}{\arcsecond} for effective spatial filtering of the background noise coming from ambient light. Our design for the PAT system aims to maintain the quantum signal within the QFOV. Lastly, to correct for the time-dependent disparity in polarization reference frames between the satellite and OGS during the satellite pass \cite{a1}, the PCS measures the polarization of the downlink beacon, which is co-aligned with the quantum signal, and actively corrects the OGS polarization axes.

\subsection{Telescope system}
From a high-level perspective, the primary purpose of the telescope system is the collection of the quantum signal. To achieve this purpose, the OGS must be able to coarsely determine the trajectory of the satellite prior to the ``Acquisition of Signal'' and be equipped with the basic capability of following the trajectory. Knowledge of the trajectory is commonly derived from the two-line element (TLE) of the satellite. The TLE provides us with the satellite's orbital information which can then be converted into a plot of celestial coordinates that is fed to the motorised tracking mount for open-loop coarse tracking. The TLE may be obtained from an online database maintained by the North American Aerospace Defense Command \cite{a10} or generated from the satellite's on board GPS telemetry data. This feature provides the OGS with basic coarse tracking capabilities based on the satellite's orbital information.

Some of the high-level requirements for the telescope system are listed below. The list is not exhaustive and is kept generic in most cases so it may be tailored more specifically based on the specific mission objectives and the specifications of the satellite and the quantum signal.

\begin{enumerate}
    \item Good optical aperture-to-cost ratio;
    \item Optimal optical efficiencies in the quantum signal and downlink beacon wavelengths (\SI{785}{\nm} and \SI{680}{\nm} respectively);
    \item Minimal optical aberrations (or aberration-free) output;
    \item Weatherised optical receiver and tracking mount;
    \item Adequate weight loading limit of the mount for optical receiver and other subsystems;
    \item Capable of slewing at a \SI{1}{\degree\per\second} or faster;\footnote{Calculated for tracking a satellite in LEO near the Zenith.}
    \item Continuous travel range of the mount in altitude and azimuth angles of \SIrange{0}{90}{\degree} and \SIrange{0}{360}{\degree} respectively; and
    \item Capable of satellite tracking based on orbital information with adequate precision and accuracy.
\end{enumerate}

As with astronomical purposes, modern large aperture optical receivers are reflecting or catadioptric telescopes for ease of production, weight, and cost reasons. It is hence common to find reflecting telescopes being utilised in fixed optical ground stations \cite{a11}. And as a result, there is an inherent optical loss due to the design, as these telescopes typically have a central obstruction. This central obstruction may differ in magnitude between manufacturers even for the same primary aperture.

Besides the basic capabilities listed above, for greater pointing and tracking accuracy, the telescope mount should come with plate solving and pointing model modules. Typical of astronomical observations, most telescope systems are equipped with plate solving and pointing model modules to help calibrate and orientate the telescope's pointing. Using the plate solving module, we would be able to accurately determine the direction in which the optical receiver is pointing from the images obtained using the receiver. By repeating this for images taken over a wider area of the sky, this would allow us to build a pointing model which could then accurately translate the motion of the mount to pointing coordinates.

\subsection{Pointing, acquisition, and tracking system}
While the telescope mount can execute basic pointing and tracking capabilities as described in the previous section, it is alone insufficient for achieving a pointing and tracking accuracy required for maintaining the quantum signal within a \SI{15}{\arcsecond} field-of-view. We need a more refined closed-loop coarse and fine tracking scheme to enhance the pointing and tracking capabilities of the OGS. In our case, we equip the OGS with two downlink beacon acquisition cameras and an uplink beacon. We base our design methodology for the PAT system on a `field-of-view' transfer concept where we first locate the satellite on a wide field-of-view beacon acquisition camera. Subsequently, the pointing error is minimized to within a smaller margin before we finally observe the quantum signal in the QFOV. We breakdown the PAT sequence into the following phases:

\begin{enumerate}
    \item Uplink beacon pointing;
    \item Open-loop coarse tracking;
    \item Closed-loop coarse tracking; and
    \item Closed-loop fine tracking.
\end{enumerate}

\textbf{Uplink beacon pointing}: In the first phase of PAT sequence, the uplink beacon is initialised and begins pointing when the satellite has risen above a threshold elevation angle. This threshold is introduced to ensure that the uplink beacon does not point into populated areas for laser safety concerns.

\textbf{Open-loop coarse tracking}: Simultaneously, the telescope mount slews to the satellite using the TLE provided to the mount control software. A wide field-of-view (WFOV) beacon acquisition camera that is side-mounted provides a visual indication of the tracking performance in this phase. The telescope mount will continue to passively track the satellite (open-loop coarse tracking) along the predicted satellite track.

\textbf{Closed-loop coarse tracking}: We first determine the pointing error accumulated in the previous phase by implementing a centroiding algorithm using the images obtained from the WFOV camera. Then, we minimize the error by communicating the calculated offset to the mount control software. Our goal is to reduce the pointing error to within the field-of-view of the narrow field-of-view (NFOV) beacon acquisition camera. In doing so, we correct the pointing error introduced due to imperfect knowledge of the satellite's orbit and any systematic tracking errors inherent in the telescope mount.

\textbf{Closed-loop fine tracking}: In this phase, we initialize closed-loop fine tracking after locating the satellite in the WFOV camera. The downlink beacon should be visible on the NFOV camera, and we can minimize any remaining pointing error by beam steering. Unlike the WFOV camera, we mount the NFOV camera on the optical receiver. A fast steering mirror (FSM) with a maximum bandwidth of \SI{600}{\hertz} is introduced into the combined beam path of the quantum signal and downlink beacon to guide the quantum signal into the QFOV. The NFOV camera provides the visual feedback required for closed-loop fine tracking. Hence, we ensure that the quantum signal remains centered in the QFOV throughout the pass, and the collection of the quantum signal is maximal.

\begin{figure*}[htbp]
    \centerline{\includegraphics[width=0.9\linewidth]{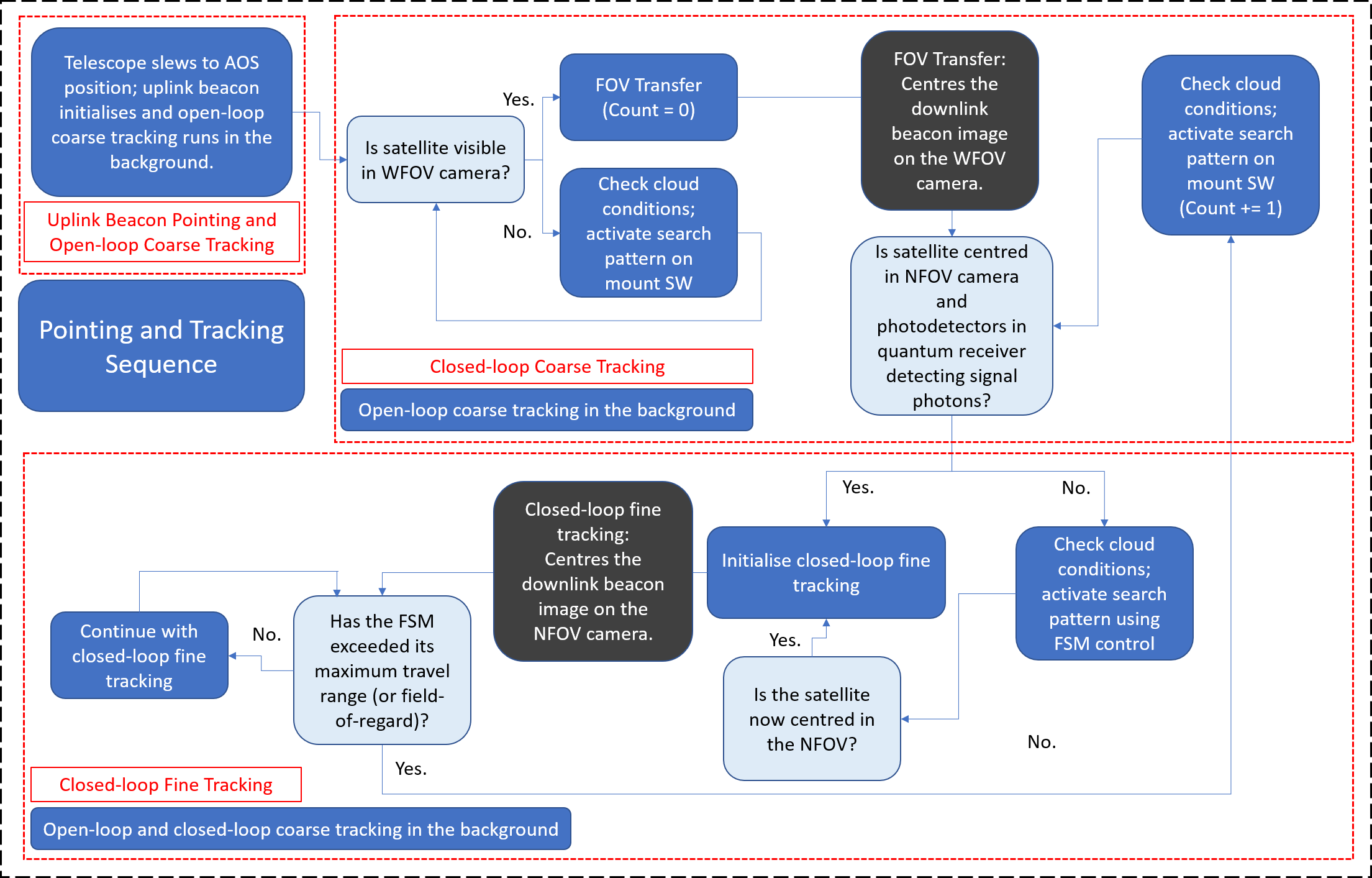}}
    \caption{Pointing, acquisition, and tracking sequence diagram.}
    \label{fig:pat-seq}
\end{figure*}

An overview of the PAT sequence may be formulated into a decision tree diagram as shown in Fig.~\ref{fig:pat-seq}.

\subsection{Polarization correction system}

\begin{figure}[htbp]
    \centerline{\includegraphics[width=0.7\linewidth]{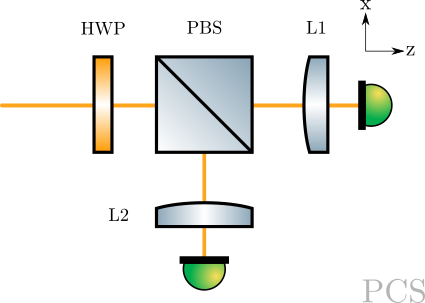}}
    \caption{Optical layout of the polarization correction system.}
    \label{fig:pcs-layout}
\end{figure}

Inherent of polarization-based entanglement QKD, we equip the OGS with a quantum receiver or a polarization analyzer (discussed in the following section). For the quantum receiver to operate effectively and to generate secret keys efficiently, the OGS must correct for dynamic offsets in the polarization reference frames between the satellite and the OGS during the pass \cite{a1}. Here, the PCS measures and determines the offsets and subsequently corrects them. For this purpose, the satellite has a polarized downlink beacon which we use as a polarization reference to the quantum signal. On the OGS, the PCS measures the polarization of the downlink beacon and calculates the visibility of the downlink beacon along two polarization axis. We execute this measurement using a basic polarimeter assembled from a pair of avalanche photodiodes, a polarizing beam splitter, and a half-wave plate (see Fig.~\ref{fig:pcs-layout}). Finally, we calculate the dynamic polarization offsets from the visibility measurement of the downlink beacon and communicate the offsets to the quantum receiver for correction.

\subsection{Quantum receiver}
While corrections to the offset between polarization reference frames at the OGS and quantum signal happen using the downlink beacon, we use the quantum signal strictly for secret key generation. As mentioned, we restrict the QFOV to \SI{15}{\arcsecond} for effective background noise suppression. Otherwise, the design of the quantum receiver unit is identical to the detection set-up in the CubeSat's science instrument, described in section \ref{sec:payload_design}. As such, we would not go into further details on the design of the quantum receiver.

\section{Conclusion}
\label{sec:conclusion}
We presented the architecture and preliminary performance of our science instrument for satellite based QKD. With a source brightness of \SI{13.6E6}{counts\per\second\per\milli\watt} and visibility of \SI{95}{\%}, we are confident that a flight model based on our instrument design is able to transmit secret keys to an optical ground station receiver with sufficient key rate. To demonstrate exactly this, we are in the process of constructing an optical ground station in Singapore. We presented the concept of operations as well as the most important subsystems of said station, which will be operational by end 2022.

\vspace{12pt}

\end{document}